\documentclass[11pt,a4paper]{article}
\pdfoutput = 1
\usepackage{jcappub}
\usepackage[utf8]{inputenc}
\usepackage{amsmath,amssymb}
\usepackage{graphicx}
\usepackage{hyperref}
\usepackage{natbib}
\usepackage{mathtools}

\author[a,b]{Joseph DeRose}
\author[b,c]{Shi-Fan Chen}
\author[a,b,c]{Martin White}
\author[d,e]{Nickolas Kokron}

\affiliation[a]{Lawrence Berkeley National Laboratory, 1 Cyclotron Road, Berkeley, CA 93720, USA}
\affiliation[b]{Berkeley Center for Cosmological Physics, Department of Physics, UC Berkeley, CA 94720, USA}
\affiliation[c]{Department of Physics, University of California,Berkeley, CA 94720}
\affiliation[d]{Department of Physics, Stanford University, 382 Via Pueblo Mall, Stanford, CA 94305, USA}
\affiliation[e]{Kavli Institute for Particle Astrophysics and Cosmology, SLAC National Accelerator Laboratory, 2575 Sand Hill Road, Menlo Park, CA 94025, USA}
\emailAdd{jderose@lbl.gov}
\emailAdd{shifan\_chen@berkeley.edu}
\emailAdd{mwhite@berkeley.edu}
\emailAdd{kokron@stanford.edu}

\title{Neural Network Acceleration of Large-scale Structure Theory Calculations}


\abstract{We make use of neural networks to accelerate the calculation of power spectra required for the analysis of galaxy clustering and weak gravitational lensing data. For modern perturbation theory codes, evaluation time for a single cosmology and redshift can take on the order of two seconds. In combination with the comparable time required to compute linear predictions using a Boltzmann solver, these calculations are the bottleneck for many contemporary large-scale structure analyses. In this work, we construct neural network-based surrogate models for Lagrangian perturbation theory (LPT) predictions of matter power spectra, real and redshift space galaxy power spectra, and galaxy--matter cross power spectra that attain $\sim 0.1\%$ (at one sigma) accuracy over a broad range of scales in a $w$CDM parameter space. The neural network surrogates can be evaluated in approximately one millisecond, a factor of 1000 times faster than the full Boltzmann code and LPT computations. In a simulated full-shape redshift space galaxy power spectrum analysis, we demonstrate that the posteriors obtained using our surrogates are accurate compared to those obtained using the full LPT model. We make our surrogate models public at \href{https://github.com/sfschen/EmulateLSS/tree/main}{https://github.com/sfschen/EmulateLSS}, so that others may take advantage of the speed gains they provide to enable rapid iteration on analysis settings, something that is essential in complex contemporary large-scale structure analyses. }

\begin{document}
\maketitle
\flushbottom


\section{Introduction}
Recent years have seen the maturation of perturbation-theory based models of large-scale structure statistics \cite{Macdonald2009,Baumann2010,Vlah2015,Vlah2016,Lewandowski2018,Chen21} with applications from redshift space distortions to synergistic combinations with $N$-body simulations for real space clustering and lensing analyses \cite{Modi2019,Kokron2021,Zennaro21,Kokron2021,hadzhiyska21}. In conjunction with the theory, a number of numerical techniques have been devised to significantly speed up the evaluation of perturbation theory integrals due to nonlinearities in gravitational clustering. With these advances, state-of-the-art perturbation theory codes \cite{classpt,pybird,Chen20} can produce fully general effective-theory predictions for galaxy clustering in any given cosmology within (roughly) two seconds.

Despite these advances, however, direct evaluation of these theory predictions remains a significant computational expense in the analysis of clustering data via Markov-chain Monte Carlo (MCMC) samplers. For full $\Lambda$CDM parameter fits in particular, the amount of time taken to generate linear power spectra and cosmological evolution parameters through Boltzmann codes like CAMB or CLASS \cite{CAMB,CLASS} typically exceeds even that spent running the perturbation codes, making the combined computational cost of performing linear power spectrum and perturbation theory calculations on the order of four seconds.

Recently, a number of authors have sought to circumvent this bottleneck by constructing fast surrogates for linear and nonlinear matter power spectra through neural networks \cite{Mancini21} as well as 2LPT and perturbation-theory/$N$-body hybrids for real space galaxy power spectra \cite{Zennaro21}. The idea of emulating such predictions through interpolating a well-sampled set of pre-computed points is not new \citep{Fendt2006,Heitmann2006,Agarwal2013}, and due to the smoothness of the dependence of galaxy clustering and weak lensing power spectra on cosmological parameters even simpler approaches such as Taylor expanding about a fiducial cosmology have been successfully implemented \cite{pybird,cataneo16,colas20,Chen21b,Osato21}.

Neural networks can be used as ``universal'' function surrogates, whose accuracy is limited primarily by the amount of data available to train them.  In this regard the combination of neural networks and perturbative models is particularly attractive, because it is possible to generate $10^5-10^6$ models to train the network at modest computational cost.  In a sense, we pay the cost of evaluating models in a given family up front, and such models can be used in many analyses.  This technique is gaining currency in cosmology, and similar approaches have been presented in refs.~\cite{Auld08,Agarwal14,Albers19,Manrique20,Kasim20,Arico21,Angulo21,Mancini21}. 

In this work, we focus on constructing surrogate models for all power spectra required for standard analyses of (real and redshift space) galaxy clustering, weak lensing, and their cross-correlation. In particular, we construct surrogates for real space galaxy auto power spectra, galaxy--matter cross power spectra and matter power spectra, $P_{gg}(k)$, $P_{gm}(k)$ and $P_{mm}(k)$ respectively, as predicted by convolutional Lagrangian effective field theory (CLEFT) \cite{Vlah2015,Vlah2016}, and hybrid effective field theory (HEFT) \cite{Modi2019, Kokron2021}. Additionally, we construct surrogates for multipoles of the redshift space galaxy auto power spectrum $P_{\ell}(k)$ as predicted by Lagrangian effective field theory \cite{Chen20,Chen21}. The surrogates that we construct have already seen their first use as applied to the 3D power spectrum and correlation function of BOSS galaxies \cite{Dawson13} and the cross-correlation between luminous red galaxies (LRGs) targeted by the Dark Energy Spectroscopic Instrument (DESI \cite{DESI}) and CMB lensing from Planck \cite{PlanckLens18} in refs.~\cite{Chen21b,White21}.

The structure of this paper is as follows. In section~\ref{sec:training_sets} we describe the models that we build surrogates for, and the methodology that we use to produce our training sets. In section~\ref{sec:methodology} we describe the combination of principal components and neural networks that we use as surrogate models, and in section~\ref{sec:results} we show that these surrogates are accurate enough for precision cosmological constraints, concluding with some final remarks in section~\ref{sec:conclusion}.

\section{Parameters and Training Sets}
\label{sec:training_sets}

We have trained surrogates for a variety of models relevant for the prediction of galaxy clustering and weak lensing statistics. The following subsections briefly describe the models for which we have built surrogates. In all cases, we sample our parameter space using a Korobov quasi-random sequence \citep{Korobov1959}, although the choice to use Korobov sequences instead of a different experimental design is insignificant given the large number of samples that we are able to produce in this work. 

All training data is generated with a publicly available \texttt{Cobaya} \cite{CobayaSoftware} likelihood\footnote{\href{https://github.com/martinjameswhite/CobayaLSS}{https://github.com/martinjameswhite/CobayaLSS}} in order to facilitate the training of surrogates for additional statistics. One particularly useful aspect of \texttt{Cobaya} that we make use of in this work is built in fast-slow parameter splitting, which allows us to generate training examples more quickly in directions of parameter space that are fast, such as galaxy bias parameters.

\begin{table*}
\begin{minipage}{\textwidth}
\caption{Parameters and priors}

\begin{center}
\begin{tabular}{|c  c  c|}
\hline
\hline
Parameter & Prior & Statistic\\  
\hline 
\multicolumn{3}{|c|}{{\bf Cosmology}} \\
$\Omega_c h^2$ &  flat ($0.08, 0.16$) & all \\ 
$\Omega_b h^2$ &  flat ($0.019, 0.024$) & all \\ 
$\log 10^{10} A_s$ &  flat ($2, 4$) & all \\
$n_s$ &  fixed (0.967) & all \\
$h$  &  flat ($0.55, 0.91$) & all \\
$\sum m_{\nu}$  &  fixed (0.06 eV) \footnote{Fixed to zero for HEFT} & all \\
\hline
\multicolumn{3}{|c|}{{\bf Galaxy Bias}} \\
$b_{1}$  & flat ($0, 3.0$) & $P_{\rm gm},  P_{\rm gg},  P_{\ell}$\\
$b_{2}$  & flat ($-5, 5$) & $P_{\rm gm},  P_{\rm gg},  P_{\ell}$\\
$b_{s}$\footnote{Fixed to zero for CLEFT}  & flat ($-20, 20$) & $P_{\rm gm},  P_{\rm gg}, P_{\ell}$ \\
$b_{k}$\footnote{HEFT only}  & flat ($-5, 5$) & $P_{\rm gm},  P_{\rm gg}$ \\
$\alpha_{0}$  & flat ($-200, 200$) & $P_{\rm gm},  P_{\rm gg}, P_{\ell}$ \\
$\alpha_{2}$  & flat ($-200, 200$) & $P_{\rm gm},  P_{\rm gg}, P_{\ell}$ \\
$R_h^3$ & flat ($-2\times 10^4, 2\times 10^4$) & $P_{\rm gg}, P_{\ell}$ \\
$R_h^3\sigma^2$  & flat ($-1\times 10^5, 1\times 10^5$) & $P_{\ell}$ \\
\hline
\end{tabular}
\end{center}
\label{tab:params}
\end{minipage}

\end{table*}
\subsection{Real Space Galaxy Power Spectra}
In order to model projected galaxy clustering and the cross-correlation between galaxy density and weak lensing statistics, we also train models for galaxy auto power spectra, $P_{\rm gg}(k)$ and galaxy--matter cross power spectra, $P_{\rm gm}(k)$. In particular, we build surrogate models for Convolutional Lagrangian Effective Field Theory (CLEFT) \citep{Vlah2016} and Hybrid Effective Field Theory (HEFT) \citep{Modi2019,Kokron2021}.

In CLEFT, the galaxy auto power spectrum and galaxy--matter cross spectrum are given by:
\begin{align}
    P_{\rm gg} &= \left(1-\frac{\alpha_{a}k^2}{2}\right)P_{\rm Z} + P_{\rm 1-loop} + b_1 P_{\rm b_1} + b_2 P_{\rm b_2} + b_1 b_2 P_{\rm b_1 b_2}  + b_1^2 P_{\rm b_1^2} + b_2^2 P_{\rm b_2^2}  + P_{\rm SN}  \\
    P_{\rm gm} &= \left(1-\frac{\alpha_{\times}k^2}{2}\right)P_{\rm Z} + P_{\rm 1-loop} + \frac{b_1}{2}P_{\rm b_1} + \frac{b_2}{2}P_{\rm b_2}\, .
\label{eq:cleft}
\end{align}
Note that we have set the shear bias ($b_s$) to zero, consistent with the findings in \citep{Modi2017}. Here, $P_{\rm Z}$ and $P_{\rm 1-loop}$ are the Zeldovich and 1-loop matter contributions, $P_{\rm SN}$ is the shot-noise contribution, $b_1$ and $b_2$ are the linear and quadratic bias parameters for the galaxy sample, and $\alpha_{\times}$ and $\alpha_{a}$ are effective field theory corrections, necessary to marginalize over small-scale physics not included in this model \citep{Vlah2015}. Further details on how each contribution is computed can be found in \citep{Vlah2016,Chen20,White21}: the power spectra can all be computed as convolutions of powers of linear power spectra times known kernels. We compute these linear spectra using the \texttt{CAMB} Boltzmann solver \cite{CAMB}. As we include a non-zero neutrino mass, we make the approximation that galaxy power spectra follow the CDM and baryon distribution \citep{Castorina15}, performing our perturbation theory calculations using $P_{\rm cb, lin}(k)$, rather than the power spectrum of all matter $P_{\rm mm, lin}(k)$.

We generate predictions at 50000 points in cosmological parameter space and 20 points in bias parameter space for each point in cosmology, for a total of $10^6$ different points in cosmology and bias space. We use a distinct Korobov sequence from which to sample our bias parameters, such that we use the $20(i)$th through the $20(i+1)$th points in the bias parameter sequence at the $i$th point in the sequence employed for the cosmological parameters. For each point in cosmology and bias parameter space, we produce 20 outputs in redshift, linearly spaced between $z=0$ and $z=2$. We use the CLEFT implementation in \texttt{velocileptors}\footnote{https://github.com/sfschen/velocileptors} \citep{Chen20} to make our predictions.

Additionally, we construct surrogates for a hybrid effective field theory (HEFT) model for $P_{\rm gg}(k)$ and $P_{\rm gm}(k)$, as implemented in \texttt{anzu}\footnote{https://github.com/kokron/anzu}\citep{Kokron2021}. The expressions for the power spectra under the HEFT model are as follows:
\begin{align}
    P_{\rm gg}(k) &= \sum_{X,Y} b_{X} b_{Y}P_{XY}(k) + P_{\rm SN} \\
    P_{\rm gm}(k) &= \sum_{X} b_{X} P_{1X}(k)\, ,
\label{eq:heft}
\end{align}
where $X, Y \in \{1, \delta, \delta^2, s^2, \nabla^2\}$. Here, $P_{11}$ is the nonlinear matter power spectrum, and $P_{1, \delta}$, $P_{1, \delta^2}$, $P_{\delta, \delta^2}$, $P_{\delta, \delta}$, $P_{\delta^2, \delta^2}$ are the nonlinear completions of $P_{\rm b_1}$, $P_{\rm b_2}$, $P_{\rm b_1 b_2}$, $P_{\rm b_1^2}$ and $P_{\rm b_2^2}$, respectively, asymptoting to the latter at low $k$. See \citep{Kokron2021} for additional implementation details. Training data is generated by sampling over cosmological and bias parameters in the same way as for CLEFT, although here we restrict the boundaries of the parameter space such that they coincide with the minimum and maximum parameter values sampled by the Aemulus simulations \citep{DeRose2018}, fixing $N_{\rm eff}=3.046$, as our HEFT model is not valid beyond those bounds.

\subsection{Redshift Space Galaxy Power Spectra}

Finally, we build surrogate models for multipoles of the redshift space galaxy power spectrum as predicted by the Lagrangian perturbation theory model described in \cite{Chen20,Chen21}. This model can be thought of as the redshift space extension of the CLEFT model discussed above, replacing real space Lagrangian displacements with their redshift space counterparts, and expanding the basis of included counterterms and stochastic terms to account for the anisotropy induced by galaxy peculiar velocities. The expression for the redshift space power spectrum under this model is as follows:
\begin{align}
    P_s(\mathbf{k}) = P_s^{\rm PT}(\mathbf{k}) + \left(\alpha_0 + \alpha_2 \mu^2\right)k^2P_{\rm Zel}(\mathbf{k}) + R_{\rm h}^3\left(1 + \sigma^2 k^2 \mu^2\right)\, ,
\label{eq:lpt_rsd}
\end{align}
where $P_s^{\rm PT}(\mathbf{k})$ is the perturbative expression for the redshift space power spectrum, containing the same bias parameters as eq.~\ref{eq:cleft}, while the second and third terms in eq.~\ref{eq:lpt_rsd} are effective field theory counterterms and stochastic terms, respectively, included to remove sensitivity to unmodeled small scale physics. $P_{\rm Zel}(\mathbf{k})$ is the first order (Zel'dovich approximation) redshift space matter power spectrum. The Alcock-Paczynski (AP) effect is included as described in Appendix C.1 of \cite{Chen21b}, and $P_s(\mathbf{k})$ is summarized by computing its monopole, quadrupole and hexadecapole moments, $P_{\ell=\{0,2,4\}}(k)$. 

The training procedure for our $P_{\ell}(k)$ surrogates is slightly different from that of the rest of the surrogates in this work. Modeling the AP effect requires us to assume fiducial distances and expansion rates, and choices for these often change from analysis to analysis making it difficult to build a surrogate model for $P_{\ell}(k)$ that is appropriate for all analyses. Additionally, nearly all RSD analyses to date have neglected the effect of the evolution of $P_{\ell}(k)$ within a redshift bin. As such, we have opted to build surrogates for a single redshift, $z=0.61$, assuming a fiducial $\Lambda$CDM cosmology with $\Omega_m=0.3$ to incorporate the AP effect, although we have made it simple to train models at different redshifts and fiducial cosmologies.

\subsection{Nonlinear Matter Power Spectrum}
Finally, we also produce a surrogate for the nonlinear matter power spectrum predicted by \texttt{CAMB} \citep{CAMB} and \texttt{Halofit} \citep{Smith2002,Takahashi2012}. The only free parameters of this model are the cosmological parameters listed in Table~\ref{tab:params}. We generate training examples at 50000 points in cosmological parameter space and 20 points in redshift for each cosmology, linearly spaced between $z=0$ and $z=2$.

\section{Surrogate Modeling Methodology}
\label{sec:methodology}
We use a combination of principal component analysis (PCA) and multi-layer perceptrons (MLP), as our surrogate models, similar to the method used in \cite{Alsing20}. In particular, our surrogate model takes the form:
\begin{align}
    \Gamma^{ab}(k, \mathbf{\Omega}) \approx \sum_{i}^{N_{\rm PC}^{ab}} \alpha_i^{ab} (\mathbf{\Omega}) \mathrm{PC}_i^{ab} (k).
\end{align}
where $\mathbf{\Omega}$ is a vector of cosmological and bias parameters, $\Gamma^{ab} (k, \mathbf{\Omega}) = \textrm{arcsinh}[\tilde{P}^{ab}(k, \mathbf{\Omega})]$, and $\tilde{P}^{ab}(k, \mathbf{\Omega})$ is a whitened version of $P^{ab}(k,\mathbf{\Omega})$, the power spectra that we are modeling. In particular, the $k$ dependence of the spectra is modeled via PCA, and the parameter dependence of the PCA coefficients, $\alpha_i^{ab}$, is modeled with an MLP. In this section we briefly outline the PCA and MLP algorithms we apply to the training sets described in section~\ref{sec:training_sets} with results described in section~\ref{sec:results}.

\subsection{Principal Component Analysis}
The first step in our surrogate modeling methodology is to construct a basis of principal components for each model. To do so, we generate a library of model predictions covering a broad range of cosmological and bias parameters and redshifts as described in section \ref{sec:training_sets}. The input data is transformed to have zero mean and unit variance, and an arcsinh function is applied to reduce the dynamic range. This data is then compressed into principal components.

Let $\mathbf{X}$ be the $N \times M$ array containing our training set, where $N$ is the number of examples in our training set and $M$ is the dimensionality of each training data vector. Then a basis of principal components can be constructed by computing the eigenvectors of the covariance matrix of each model:

\begin{align}
    \mathbf{C} &= \mathbf{X}^{\rm T} \mathbf{X},\\
    &= \mathbf{W}\mathbf{\Lambda}\mathbf{W}^{\rm T},\nonumber
\end{align}
where the rows of $\mathbf{W}$ are the eigenvectors, i.e., the principal components in question and $\mathbf{\Lambda}$ is a diagonal matrix of the eigenvalues, which are equal to the variance of the model component described by each eigenvector.

Depending on our application, we vary the number of principal components employed in our surrogate model to achieve our desired accuracy. The coefficients of these principal components are then predicted by a MLP. As our training data is virtually noiseless (modulo numerical noise), we are not penalized for choosing $N_{\rm PC}$ to be larger than necessary. As such, we choose the number of principal components for each application so that
\begin{equation}
\label{eq:n_pc}
    1-{\sum_{i<N_{\rm PC}} \Lambda_{i}^{1/2} / \mathrm{Tr}(\mathbf{\Lambda}^{1/2})} < 10^{-4}\, .
\end{equation}
This ensures that the principal component representation of our data is not close to the limiting factor in the accuracy of our surrogate models.

\begin{figure}
	\includegraphics[width=\columnwidth]{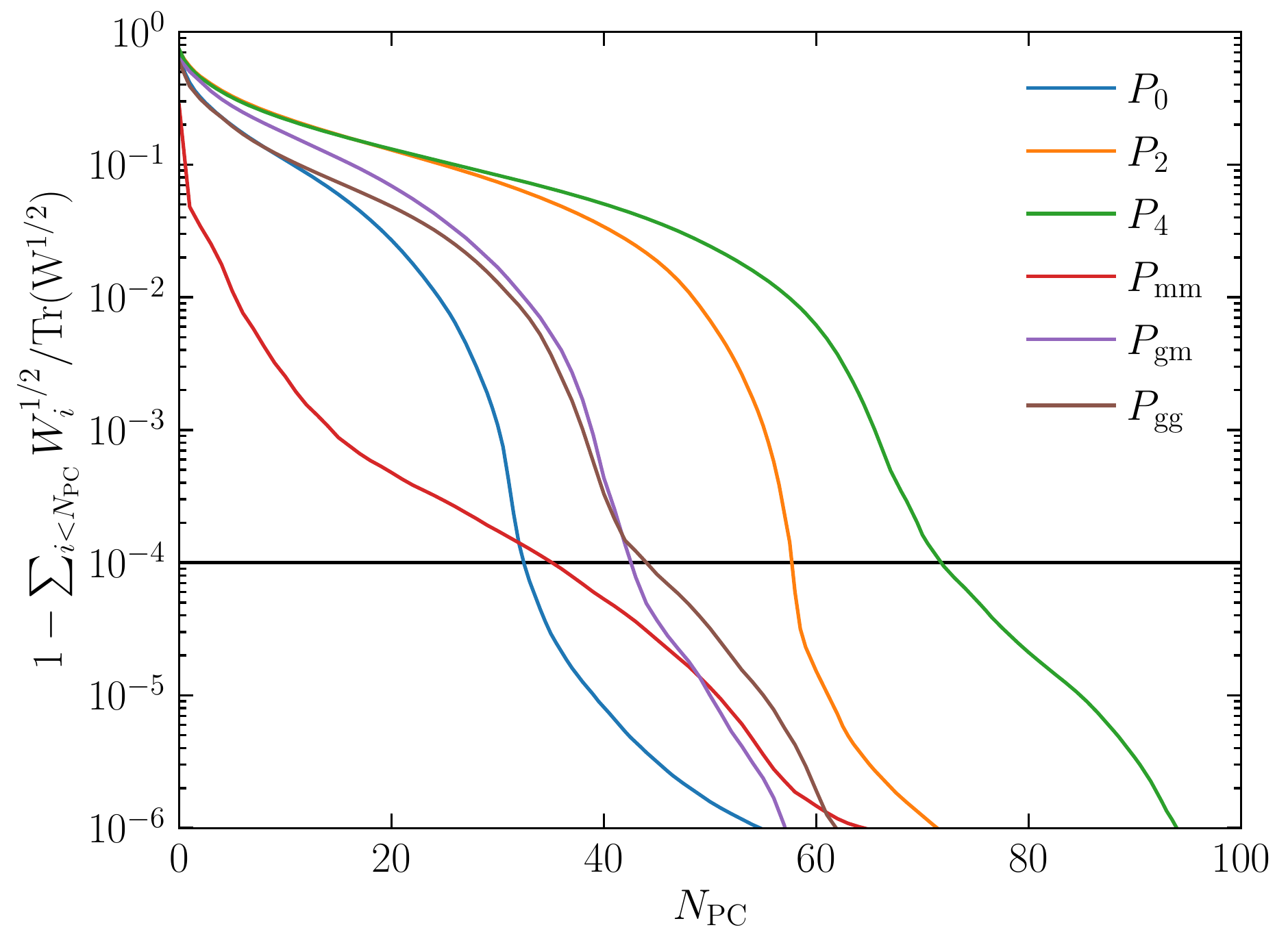}
    \caption{For each statistic we wish to model, we choose the number of principal components employed in our surrogate model by determining the minimum number of components required to satisfy the inequality in equation~\ref{eq:n_pc}. This figure shows the fractional error induced by truncating our PCA representation after $N_{\rm PC}$ principal components for the six statistics that we focus on in this work. For the real space statistics in this figure, we use the CLEFT model, but we find similar results for HEFT.}
    \label{fig:pca}
\end{figure} 

\subsection{Network Architecture and Optimization}
We use a fully connected MLP architecture with 4 hidden layers, each with 128 neurons and an activation function given by:
\begin{equation}
    a(\mathbf{x}) = \left[ \mathbf{\gamma} + \frac{1-\mathbf{\gamma}}{1+\exp(-\mathbf{\beta}\odot\mathbf{x})} \right] \odot \mathbf{x}
\end{equation}
where $\odot$ indicates element-wise multiplication, following \cite{Alsing20}. The MLP predicts $N_{\rm PC}$ values, which are then multiplied by their respective principal component vectors and summed in order to compute a mean squared error loss in the whitened training data space. We selected this architecture by performing a hyper-parameter optimization over the number of layers and neurons per layer. The optimization was performed using $P_{4}(k)$ as training data, as this statistic consistently produced the largest residuals. The resulting architecture performs well for all other statistics considered in this work.

We use one MLP per statistic, optimizing the MLP parameters using \texttt{adam} \citep{adam} and a mean squared error loss function. The inputs to our network are cosmological and bias parameters, as well as redshift where appropriate, transformed to have zero mean and unit variance. We begin the optimization process using a learning rate of $\eta=10^{-2}$, and we decrease this by an order of magnitude every 30 training epochs until we reach $\eta=10^{-6}$. We use a batch size that decreases proportionally to our learning rate, starting out by splitting our training data into $320$ batches, and halving the batch size every time we decrease the learning rate.

\section{Results}
Having described how we construct our surrogate models, we now discuss their accuracy, demonstrate their utility in a realistic analysis, and compare them to a common alternative in Taylor series expansions.

\begin{figure}
	\includegraphics[width=0.5\columnwidth]{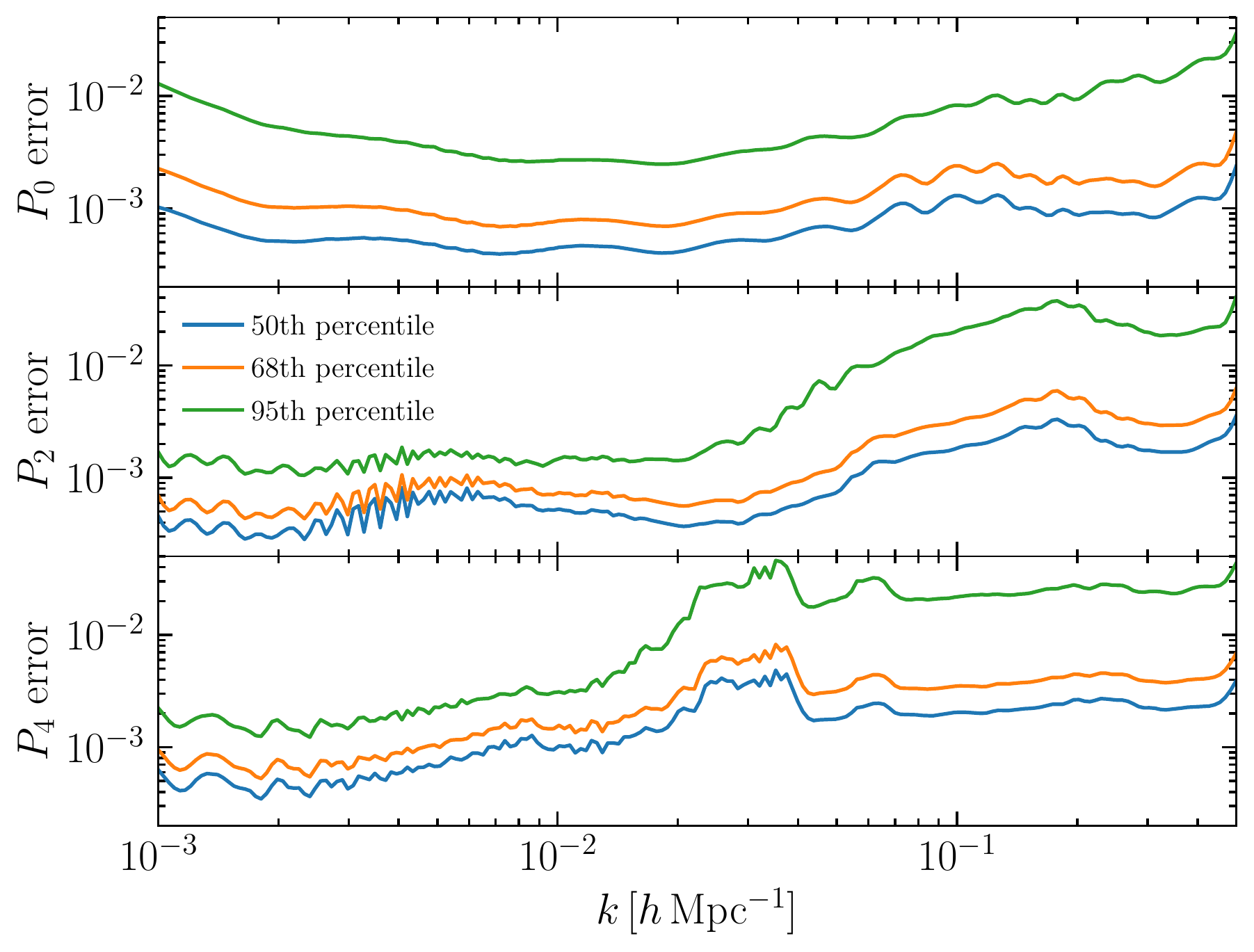}
	\includegraphics[width=0.51\columnwidth]{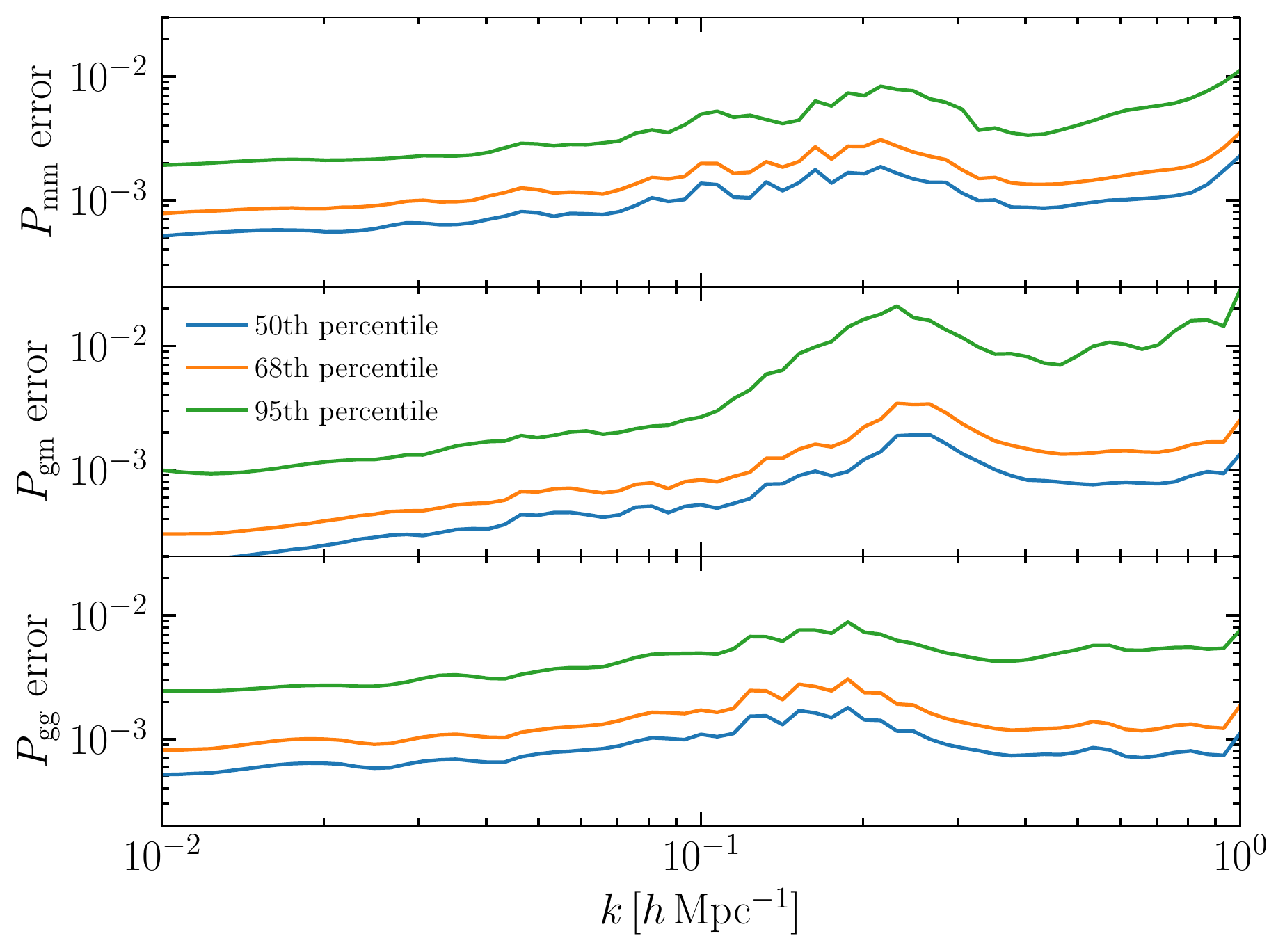}
    \caption{Fractional residuals of our surrogates when comparing to a test set drawn from the same prior bounds as the training set, but held out of the MLP optimization procedure. We display the median absolute deviation (MAD), 68th percentile and 95th percentile residuals as blue, orange and green lines respectively for each statistic. The RSD statistics are the most difficult to construct surrogates for, particularly $P_{2}$ and $P_{4}$, as they have a large dynamic range due to the counterterms and stochastic terms used in the EFT model that we employ. The residuals are made more significant by the zero crossings of $P_2$ and $P_4$, where the denominator of the fractional error becomes small. For the real space spectra, we display the residuals for our CLEFT surrogates. Our HEFT surrogates achieve very comparable levels of accuracy.}
    \label{fig:test_residuals}
\end{figure}

Figure~\ref{fig:test_residuals} shows the mean, 68th percentile and 95th percentile fractional residuals for each of our surrogate models as a function of wave number. These residuals are computed from a set of 10000 test spectra that are held out of the training process. For all surrogates, the mean residuals are frequently below $0.1\%$, with the $1\sigma$ residuals approximately $20-50\%$ larger than the mean. The $2\sigma$ residuals approach and sometimes exceed $1\%$ at high $k$, where the model predictions vary significantly more as a function of the nonlinear parameters included in the models. The redshift space power spectra are the most difficult to construct surrogates for, largely due to their larger dynamic range driven by the broad priors we allow on the bias parameters, counterterms, and stochastic terms.

Our surrogate models each take approximately one ms to evaluate, which is a factor of more than 1000 times faster than a call to \texttt{CAMB} and \texttt{velocileptors}, which takes approximately 4 seconds in combination. This means that the computational cost of generating the training data for the surrogate models is amortized with just one MCMC, assuming the chain requires approximately one million model evaluations.

\subsection{Fitting the PT Challenge}
\label{sec:pt_challenge}
In order to demonstrate the accuracy of our surrogate models in a realistic setting, we perform an MCMC analysis on the PT challenge simulations \citep{Nishimichi20} and compare posteriors obtained with surrogate models to those obtained with the exact models. The PT challenge simulations are a suite of 10 $N$-body simulations designed to test large-scale, full-shape RSD analyses. Each simulation has a side length of $3840\, h^{-1}\,\rm Mpc$ and $3072^3$ particles. The simulations are populated with a halo occupation distribution in order to roughly reproduce the large scale BOSS RSD power spectrum multipoles. The multipoles measured from these simulations have a significantly greater precision than any currently realized measurements, and thus provide a useful testing ground for analytic models. In this analysis, we focus on the $z=0.61$ measurements using the provided covariance appropriate for the volume of the BOSS survey \cite{Dawson16}\footnote{Measurements and covariance were obtained from https://www2.yukawa.kyoto-u.ac.jp/~takahiro.nishimichi/data/PTchallenge/}.

For the purposes of this test we fix $\sum m_{\nu}=0$, as opposed to what is described in Sec.~\ref{sec:training_sets}, as the PT challenge simulations assume massless neutrinos. This requires retraining the surrogate models for $P_{\ell}(k)$, but the change in accuracy when doing so is negligible. We assume priors for bias parameters as listed in Table~\ref{tab:params}, but alter the priors on cosmological parameters, setting flat priors of $0.2\le \Omega_m \le 0.4$, $50\le H_0 \le 80$ and $1.61\le \log 10^{10} A_s \le 3.91$, fixing $n_s=0.967$ and $\Omega_{b}=f_b \Omega_m$, where $f_b=0.1571$, consistent with the information provided to us by the conveners of the PT challenge. We fit the monopole, quadrupole, and hexadecapole to $k=0.12\, h\, \rm Mpc^{-1}$.

\begin{figure}
	\includegraphics[width=\columnwidth]{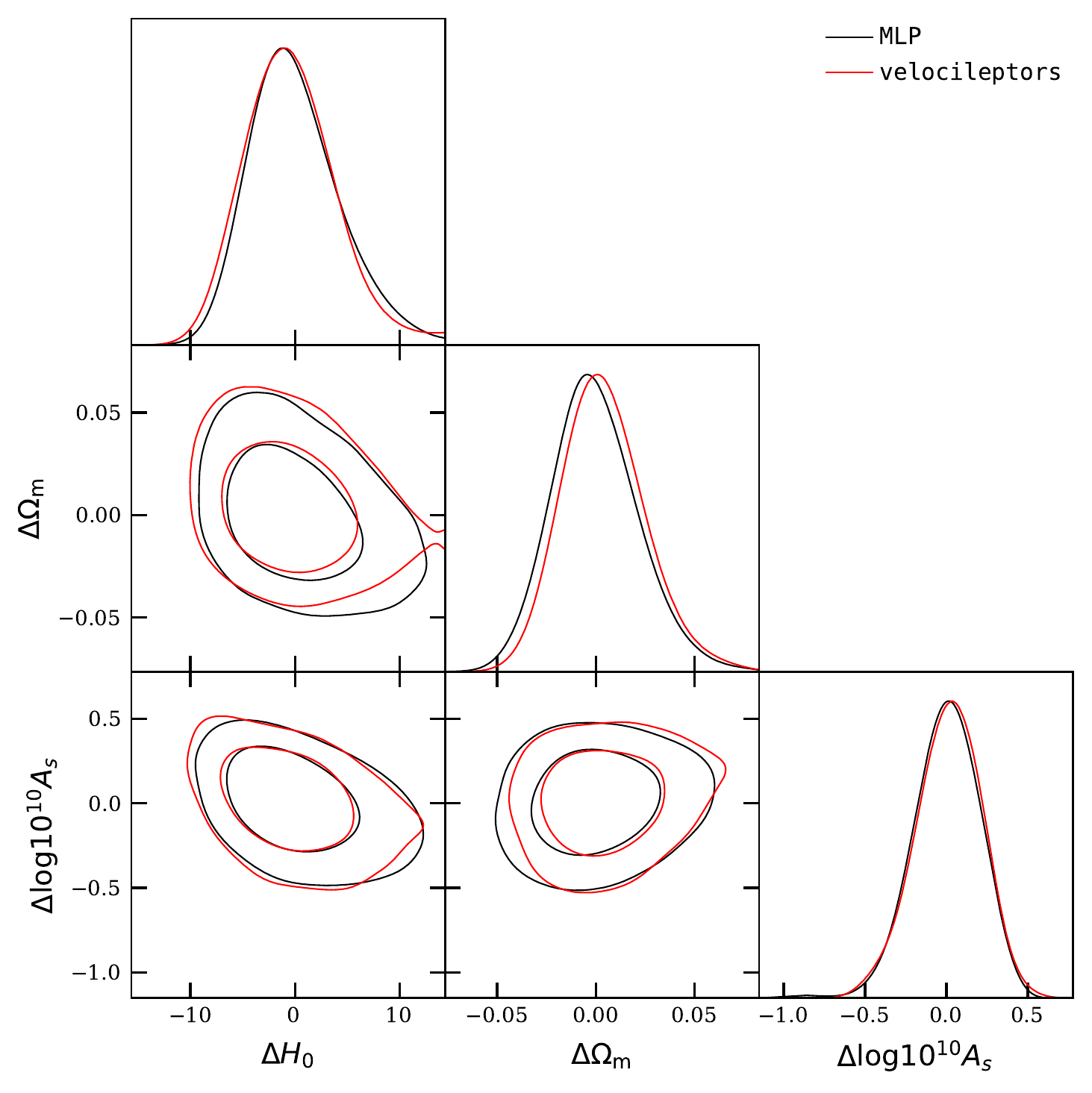}
    \caption{Constraints on the three cosmological parameters left free when fitting the PT challenge RSD multipoles. The black contours are obtained with our surrogate models, while the red use \texttt{velocileptors}. Contours are shifted to be centered at 0, so as to not reveal the input cosmology of the PT challenge simulations. The agreement is excellent.} 
    \label{fig:pt_challenge}
\end{figure}

Figure~\ref{fig:pt_challenge} compares the cosmological constraints obtained using our surrogate models for $P_{\ell}(k)$ to those obtained using the standard \texttt{velocileptors} predictions. We have shifted the posteriors to be centered at 0 to avoid unblinding the input PT challenge cosmology. The agreement between the two posteriors is excellent, with shifts in the mean parameter constraints of $\leq 0.15\sigma$, and errors agreeing at better than the $10\%$ level for all parameters.

\subsection{Comparison with Taylor Expansions}
\label{sec:taylor_expansion}

Finally, we compare our MLP surrogate model to a Taylor series expansion, a common alternative for approximating EFT models for redshift space power spectra \cite{Chen21b, cataneo16}. To set up the Taylor series at N$^{\rm th}$ order we compute derivatives via finite difference on a $(2N+1)^3$ grid in $\Omega_m, h, \ln(10^{10}A_s)$ using the \texttt{FinDiff}\footnote{ \url{https://findiff.readthedocs.io/en/latest/} } package. Specifically, since perturbation theory predictions for the redshift space multipoles given a set of cosmological parameters $\Theta$ are given by inhomogeneous quadratic polynomials in the bias and EFT parameters we can write
\begin{equation*}
    P_\ell(k|\Theta) = \sum_{ij} \beta_i \beta_j \mathcal{M}_{ij}(\Theta), \quad \beta = (1, b_1, b_2, b_s, \alpha_0, \alpha_2, R_{h}^3, \sigma^2R_{h}^3).
\end{equation*}
Since the matrix $\mathcal{M}_{ij}$ is a smooth function of the cosmological parameters $\Theta$ we can expand it as a Taylor series, such that the full expression is
\begin{equation}
    P_\ell(k|
    \Theta) = \sum_{ij} \beta_i \beta_j \sum_{M=0}^N \frac{1}{M!}(\Theta - \Theta_0)_{a_1} \cdots (\Theta - \Theta_0)_{a_M}
    \ \left. \partial_{a_1}\cdots \partial_{a_M} \mathcal{M}_{ij}\right|_{\Theta_0} .
\end{equation}
Note that in principle one can concatenate $\beta$ and $\Theta$ into one ``super'' parameter vector to Taylor expand in, but that the mathematical structure of perturbation theory limits the number of polynomial coefficients in the bias parameters, similarly to how the smoothness of the multipoles in $\Theta$ allows us to use a low-order Taylor series compared to a more generic parametrization. Comparing a Taylor series so-constructed to the more generic MLP surrogate described in this paper is the goal of this subsection.

\begin{figure}
	\includegraphics[width=\columnwidth]{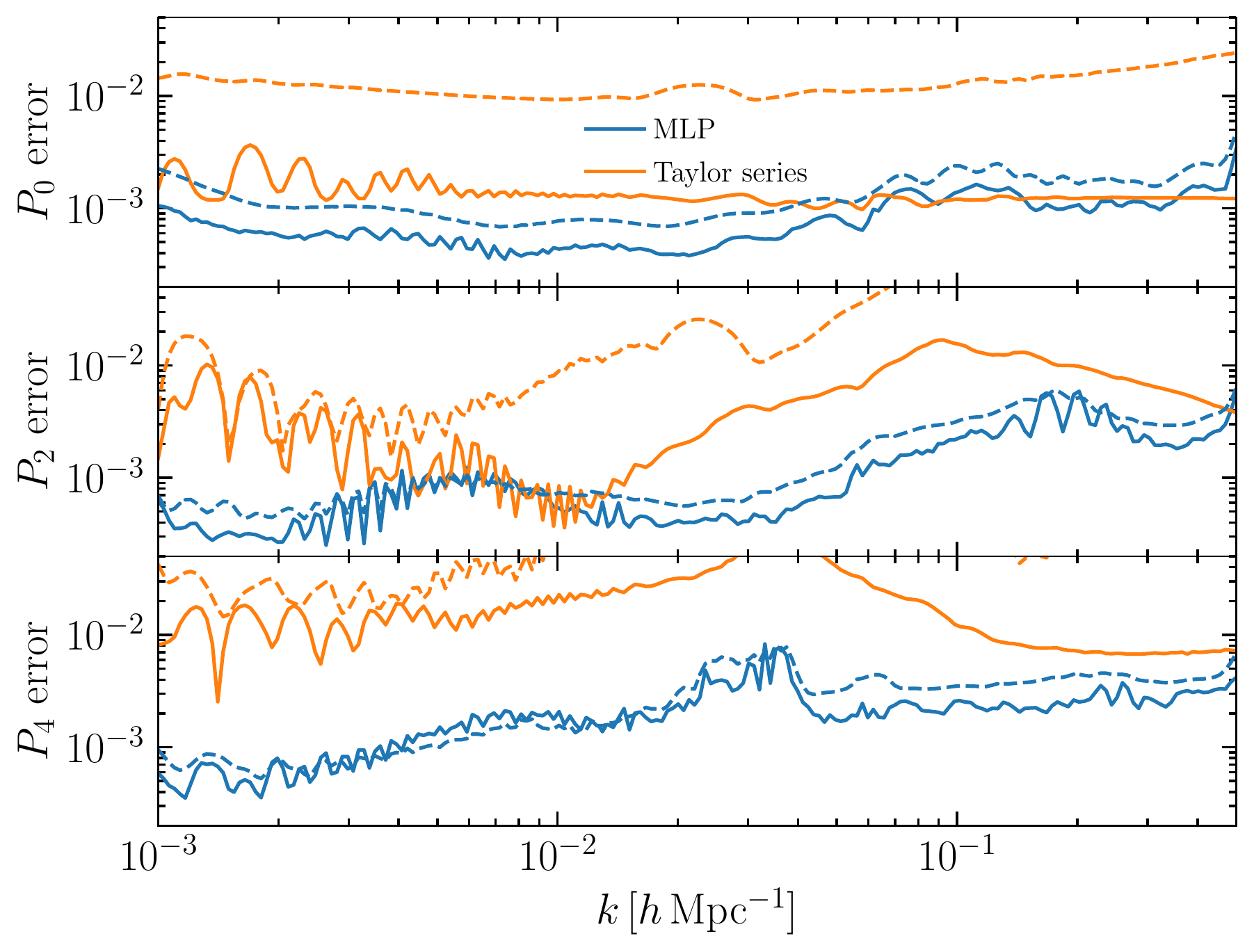}
    \caption{Comparison of our MLP surrogate model to a Taylor expansion for LPT predictions of $P_{\ell}(k)$.  68th percentile fractional residuals of our MLP (blue) and Taylor series (orange) surrogate models when comparing to our full test set (dashed) or a test set restricted to a three sigma region in cosmological parameter space around the best fit cosmology from \cite{Planck2018} (solid). In general, the MLP outperforms the Taylor series, but the surrogates achieve more comparable levels of accuracy in the restricted test set, where accuracy is most important given our prior knowledge of cosmological parameters. The Taylor series surrogate is a factor of $\mathcal{O}(10^3)$ times less expensive to generate.}
    \label{fig:taylor_comparison}
\end{figure}

Figure~\ref{fig:taylor_comparison} compares the accuracy of our MLP model to that of the Taylor series expansion with $N=4$. In particular, we compute the 68th percentile fractional error when comparing to a test set spanning a $3\sigma$ parameter range around the Planck 2018 best fit cosmology\cite{Planck2018}. The MLP performs significantly better in general, though the Taylor series often achieves errors significantly below one percent, especially for the monopole and quadrupole, where measurements are most precise. The dashed lines show the same quantities, but computed over the full parameter range that our MLP is trained on. The MLP performs similarly well in this expanded parameter range, while (unsurprisingly) the Taylor series fares significantly worse with errors often exceeding one percent.  Reducing the error would require us to produce multiple, overlapping Taylor expansions with different central points.  This increases the complexity of the method and introduces more hyperparameters to optimize.

\label{sec:results}

\label{sec:results}

\section{Conclusion}
\label{sec:conclusion}
In this work we have combined principal component analysis (PCA) and multi-layer perceptrons (MLP) to construct surrogate models for commonly used ingredients in galaxy clustering and weak lensing analyses. In particular, we have constructed surrogates for real space galaxy--galaxy, galaxy--matter, and matter--matter power spectra, and multipoles of the redshift space galaxy auto power spectrum as described in section~\ref{sec:training_sets}. We use CLEFT \citep{Vlah2016} and HEFT \citep{Modi2019,Kokron2021} to predict the galaxy--galaxy and galaxy--matter power spectra, \texttt{halofit} to predict the nonlinear matter power spectrum \citep{Smith2002, Takahashi2012}, and the full LPT model discussed in \cite{Chen21} to predict the redshift space galaxy power spectra. We train these surrogates over a wide cosmology and bias parameter space to facilitate a broad range of analyses. In section~\ref{sec:results}, we show that these surrogates achieve mean fractional errors that are at or below $0.1\%$ for most scales. We demonstrate the accuracy of our surrogate models by performing a simulated analysis on redshift space power spectrum multipoles measured from the PT challenge boxes \citep{Nishimichi20} in section~\ref{sec:pt_challenge}. Finally, we compare to a Taylor series expansion of redshift space power spectrum multipoles, demonstrating that our surrogate models achieve significantly better accuracy than these commonly used alternatives. We make our surrogate models public at \href{https://github.com/sfschen/EmulateLSS/tree/main}{github.com/sfschen/EmulateLSS}, in the hope that they will be widely applicable for ongoing cosmological analyses. 

The computational expense of parameter estimation is rapidly becoming a limiting factor in analyses of large-scale structure. This is largely due to the increasing complexity of models driven by the stringent accuracy requirements imposed by growing measurement precision. As such, it is important to find ways to expedite the parameter estimation step of analyses. In this work, we have focused on doing this by significantly speeding up the prediction of power spectra by constructing PCA+MLP surrogate models. One could also consider building surrogates for the final observables such as angular power spectra, correlation functions, or window-convolved power spectra, but doing so involves additional tracer and survey specific information that would require building new surrogates for each analysis. This may be beneficial in the case where projection or window convolution operations make up a significant portion of a single likelihood evaluation, e.g., if the Limber approximation is not applicable\citep{Krause2021}. Acceleration of theory calculations in the manner presented in this work will significantly expedite complex contemporary cosmology analyses, saving countless hours of waiting for MCMCs to converge, and empowering scientists to perform more robust inference.

\section*{Acknowledgments}
J.D.~is supported by the Chamberlain Fellowship at Lawrence Berkeley National Laboratory.
S.C and M.W.~are supported by the DOE and the NSF. N.K.~is supported by the Gerald J. Lieberman fellowship.
We acknowledge the use of \texttt{Cobaya} \cite{CobayaSoftware, Torrado21}, \texttt{GetDist} \cite{Lewis19} and \texttt{CAMB} \cite{Lewis00} and thank their authors for making these products public.
This research has made use of NASA's Astrophysics Data System and the arXiv preprint server.
This research used resources of the National Energy Research Scientific Computing Center (NERSC), a U.S. Department of Energy Office of Science User Facility operated under Contract No.\ DE-AC02-05CH11231.

\appendix


\bibliographystyle{JHEP}
\bibliography{main}
\end{document}